\documentclass[pra,aps]{revtex4}
\usepackage{amssymb}
\usepackage{amsmath}
\usepackage{epsfig}

\usepackage{colordvi}
\usepackage{graphics}
\usepackage{rotating}

\begin{document}

\title{Time and Frequency description of Optical Pulses}

\date{\today}

\author{Ludmi{\l}a Praxmeyer$^{1}$ and
Krzysztof W\'odkiewicz $^{1,}\, ^{2}$ } \affiliation{$^{1}$
Instytut Fizyki Teoretycznej, Uniwersytet Warszawski, Warszawa
00--681, Poland \\%
$^{1}$Department of Physics and Astronomy,
University of New
Mexico, Albuquerque NM 87131\\%
}

\begin{abstract}
The connection between the time-dependent physical spectrum of
light and  the phase space overlap of Wigner functions is
investigated for optical pulses. Time and frequency properties of
optical pulses with chirp are analyzed using the  phase space
Wigner  and Ambiguity distribution functions. It is shown that
optical pulses can exhibit interesting phenomena, very much
reminiscent of quantum mechanical interference, quantum
entanglement  of wave packets,  and quantum sub-Planck structures
of the time and frequency phase space.
\end{abstract}

\maketitle

\section{Introduction}

In a paper published in 1977, Professor Eberly and one of the
present authors (KW), have introduced the time-dependent physical
spectrum of light described by non-stationary  random electric
fields \cite{jhekw1977}. The definition of such an operational
spectrum required, as an essential step, a frequency tunable
filter, that allows the resolution of the frequency components. If
one restricts the result of \cite{jhekw1977} to deterministic
fields only,  the time-dependent physical spectrum takes the
following form
\begin{equation}
\label{tds}
G(t,\omega)=\biggl|\int ds\, {
\mathcal{J}(t-s;\Gamma)E(s)e^{i\omega s}}\biggl|^2\,,
\end{equation}
where $E(t)$ is the positive frequency part of the detected
electric field, and the  spectral properties of the filter are
represented by the response function
$\mathcal{J}(t;\Gamma)e^{i\omega t}$. This function   depends on
the setting frequency $\omega$, and
 the response  amplitude is characterized  by a
bandwidth $\Gamma$ of the filter. If one rewrites the
time-dependent spectrum formula (\ref{tds}) in the following form
\begin{equation}\label{wavelet}
 G(t,\Gamma)=\biggl|\int ds\, {
\mathcal{J}\bigl(\frac{t-s}{\Gamma}\bigr)E(s)e^{i\omega s}}\biggl|^2\,,
\end{equation}
and look at  (\ref{tds}) as a function of time and bandwidth only,
we recognize in this expression the modulus square of the wavelet
transformation of the signal field, where $t$ is the time shift
and $\Gamma$ is the scale \cite{mallat1999}.

Several years after the introduction of the time-dependent
spectrum, similar expressions to  (\ref{tds}) have been derived
and applied to the reconstruction of the amplitude and  phase of
optical pulses. If one selects $\mathcal{J}(t)=|E(t)|^2$  or
$\mathcal{J}(t)= E(t)$, the recorded time-dependent spectrum
corresponds to  the frequency resolved optical gating (FROG)
\cite{frog1993}, or to the second harmonic  frequency resolved
optical gating (SHG FROG) \cite{frog1994}. Methods based on FROG
have become powerful tools in the investigations of femtosecond
pulses.

The remarkably simple expression (\ref{tds}), hides a very
interesting  phase space structure of the operational spectrum in
terms of a time and frequency distribution overlap. It has been
shown \cite{kbkw1982}, that the time-dependent spectrum can be
visualized as a time and frequency convolution of two   Wigner
distribution functions
\begin{equation}\label{tdsWig}
 G(t,\omega)=2\pi \int dt^{\prime} \int d\omega^{\prime}\,
 W_{J}(t^{\prime},\omega^{\prime})
 W_{E}(t^{\prime}+t,\omega^{\prime}+\omega)\,.
\end{equation}
In this expression  the first Wigner function corresponds to   the
time and frequency distribution of the filter, while the second
Wigner function describes the  phase space properties of the
measured electric field. It is perhaps worth mentioning, that a
quantum version  of  the expression (\ref{tdsWig}), can be applied
in quantum mechanics, to describe joint operational   measurements
of position and momentum \cite{kw1984a}.

In order to honor Professor Eberly's contributions to the
development of the time-dependent spectrum and Quantum Optics,
this paper will investigate  time and frequency properties of
optical pulses using the concept of the phase space Wigner
distribution. We will show that optical pulses can exhibit
interesting phase space structures very much reminiscent of
quantum mechanical interference \cite{cgplk2005}, quantum
entanglement of wave packets \cite{laweberly2004}, and quantum
sub-Planck structures \cite{zurek2001}.

This paper is organized in the following way. Section II is
devoted to the definition and elementary properties of the time
and frequency Wigner  and Ambiguity distribution functions.
Section III contains a detailed description of chirped pulses
using a time and frequency phase space. We show that  time and
frequency correlations of chirped pulses  have a formal analogy to
entanglement of wave packets. The strength of these correlations
is investigated using the Schmidt decomposition. In Section IV we
analyze the phase space properties of a linear superposition of
two chirped pulses. In Section V we discuss the connection of
sub-Fourier phase space structures with pulse overlaps. Finally
some concluding results are presented.

\section{Time and Frequency Phase Space}

\subsection{The Wigner Function}

We shall investigate time and  frequency properties of optical
pulses  using a phase space distribution function. Such a function
has been originally introduced   by Wigner in 1932 and applied to
quantum mechanics \cite{wigner1932}. In the area of signal
processing the same distribution function has been used by Ville
in 1948 \cite{ville1948}. The time and frequency Wigner
distribution function corresponding to the field envelop $E(t)$ is
defined as
\begin{equation}
\label{wigfunc} W_{E} (t,\omega) = \int \frac{d s}{2\pi} \;
E^{\ast}\left( t+\frac{s}{2} \right) \; e^{i\omega s} \; E\left(
t-\frac{s}{2} \right)\,.
\end{equation}
It is well known that the Wigner function can be used as a time
and frequency distribution of the pulse, but cannot be guaranteed
to be positive for all fields. There is an extensive literature
devoted to the properties and applications of the Wigner function
in quantum mechanics \cite{schleich2001} and classical optics
\cite{cohen1995}. Below we present only the most relevant
properties of the Wigner distribution needed for the purpose of
this paper.

The frequency integration  of the Wigner function yields  the
temporal instantaneous intensity
\begin{equation}\label{tmarg}
\int d\omega \, W_{E}(t,\omega)= I(t) =  E^{\ast}(t)E(t)\,.
\end{equation}
The corresponding  time integration of this distribution leads to
the power spectrum of the optical pulse
\begin{equation}\label{omegamarg}
\int d t\,  W_{E}(t,\omega)= P(\omega) =
\frac{1}{2\pi}\tilde{E}^{\ast}(\omega)\tilde{E}(\omega)\,.
\end{equation}
In this formula the expression $\tilde{E}(\omega) = \int dt
e^{-i\omega t} E(t)$  is the Fourier transform of the pulse. From
these definitions we see that that the Wigner function is
normalized to the total power/energy of the pulse
\begin{equation}\label{wignormal}
 \int d t \int d\omega\, W_{E}(t,\omega)= \int dt\, I(t)= \int
 d\omega\,
 P(\omega)\,,
\end{equation}
where the last equality follows  from the  Parseval theorem for
the Fourier transforms. An important result that we shall use in
the following sections is the overlap relation for two Wigner
functions
\begin{equation}\label{overlap}
|\langle E_1|E_2\rangle|^2= \left|\int dt E^{\ast}_{1}(t)E_{2}(t)\right|^2=
2\pi \int dt \int d\omega\, W_{E_1}(t,\omega) W_{E_2}{(}t,\omega)\,.
\end{equation}
This formula indicates that  the case a zero overlap of two pulses
is impossible to achieve with positive Wigner functions.

Using the Wigner function as a weighting distribution one can
characterize the  properties of the optical pulse  in the form of
the following statistical moments of time and frequency
\begin{equation}\label{twmoments}
\langle t^n\omega^m\rangle =\frac{\int dt \int d\omega\,
t^n\omega^m \,W_{E}(t,\omega) }
   { \int dt \int d\omega\, W_E(t,\omega) }\,.
\end{equation}

\subsection{The Ambiguity function}

A different way of looking at the time-frequency correlations
(\ref{twmoments}) is to use the Ambiguity function, which is a
two-dimensional Fourier transform of the Wigner function
\begin{eqnarray}
\label{ambfunc} A_{E}(T,\Omega)& =&  \int dt \int d\omega\,
 e^{i\Omega t+i \omega T} W_{E}(t,\omega)      \nonumber\\
&=& \int dt \; E^{\ast}\left( t-\frac{T}{2}\right) \; e^{i\Omega t
} \; E\left( t+\frac{T}{2}\right)\,.
\end{eqnarray}
From this  definition it follows that the Ambiguity function can
be written as
\begin{equation}
\label{genfunc}
 A_{E}(T,\Omega)=A_{E}(0,0) \langle e^{i\Omega t+i \omega
 T}\rangle\,,
\end{equation}
where $ A_{E}(0,0) = \int dt \int d\omega\, W(t,\omega)$ is a
normalization constant. The formula (\ref{genfunc}) can be used as
a moment generating function. The   time and frequency statistical
moments (\ref{twmoments}) can be calculated from the Ambiguity
function using the formula
\begin{equation}\label{tfreqdispdef}
\langle t^n\omega^m\rangle = \frac{(-i)^{n+m}}{A_{E}(0,0)}
\frac{d^{n}}{d\Omega^n}\frac{d^{m}}{dT^m}
 A_{E}(T,\Omega)\biggl|_{T=\Omega=0}\,.
\end{equation}

\subsection{ABCD optics of optical pulses}

It is known from classical optics  that for linear optical devices
one can use the $ABCD$ transformation of geometrical ray
displacement and slope \cite{pmjhe}. We will use this approach to
describe arbitrary linear transformations of time and frequency
given by
\begin{equation}\label{ABCDm}
 \left(%
\begin{array}{c}
 t^{\prime} \\
  \omega^{\prime} \\
\end{array}%
\right)=   \left(%
\begin{array}{cc}
  A & B \\
  C & D \\
\end{array}%
\right)
\left(%
\begin{array}{c}
 t \\
  \omega \\
\end{array}%
\right)\,.
\end{equation}
This transformation is canonical if it preserves the normalization
of the Wigner function

\begin{equation}\label{canonical}
 \int dt \int d\omega
\,W(At+B\omega,Ct+D\omega) = \int dt \int d\omega \,W(t,\omega)\,.
\end{equation}
This condition is satisfied if
\begin{equation}\label{normal}
    \det \left(%
\begin{array}{cc}
  A & B \\
  C & D \\
\end{array}%
\right)=1\,.
\end{equation}
The $ABCD$ transformation of the Wigner function generates the
following transformation of the Ambiguity function
\begin{equation}
W(At+B\omega,Ct+D\omega)\Rightarrow
A_{E}(At-B\omega,D\omega-Ct)\,.
\end{equation}

\section{Time and frequency description of chirped pulses}

\subsection{General properties of chirped pulses}
 Let us write the electric field in the form of a real
envelop and phase:
\begin{equation}\label{envelop}
E(t)=\mathcal{E}(t) e^{-i\varphi(t)}\,.
\end{equation}
In order to calculating the Wigner function for such a pulse we
use a linear approximation for the phase:
$\varphi(t+s/2)-\varphi(t-s/2)\simeq \dot{\varphi} s$, and as a
result we  obtain
\begin{equation}\label{wphase}
    W_{E}(t,\omega)= W_{\mathcal{E}}(t,\omega+\dot{\varphi})\,,
\end{equation}
where $W_{\mathcal{E}}(t,\omega)$ is the Wigner function  of the
real envelop $\mathcal{E}(t)$. The corresponding formula for the
Ambiguity function is
\begin{equation}
    A_{E}(t,\omega)=
    A_{\mathcal{E}}(t,\omega-\dot{\varphi})\,.
\end{equation}
The instantaneous pulse frequency $\omega(t)$ of a chirped pulse
$E(t)=\mathcal{E}(t) e^{i\omega(t)}$, can be defined as
\begin{equation}
\omega(t) =\frac{ \int d\omega\, \omega \,W_{E}(t,\omega) }
   {  \int d\omega\, W_{E}(t,\omega) }=
   \langle \omega\rangle_{\mathcal{E}} -\dot{\varphi}\,,
\end{equation}
and the corresponding square of instantaneous pulse frequency is
\begin{equation}
\omega^2(t) =\frac{ \int d\omega\, \omega^2 \,W_{E}(t,\omega) }
   {  \int d\omega\, W_{E}(t,\omega) }=
   \langle \omega^2\rangle_{\mathcal{E}} -2\dot{\varphi}\langle \omega\rangle_{\mathcal{E}}
   + \dot{\varphi}^2\,,
\end{equation}
where $\langle \omega\rangle_{\mathcal{E}}$ and $\langle
\omega^2\rangle_{\mathcal{E}}$ are the frequency two moments,
calculated with respect to the unchirped amplitude $\mathcal{E} $.
From these relations we conclude that the dispersion of the
instantaneous pulse frequency is
\begin{equation}\label{dispuls}
  (\Delta \omega(t))^2=(\Delta \omega)_{\mathcal{E}}^2 +
  (\Delta\dot{\varphi}(t))^2\,.
\end{equation}

\subsection{Gaussian optical pulses with linear chirp}

As an example of the general envelop (\ref{envelop}), we will
consider a single pulse with a linear chirp and Gaussian envelop
function of the form
\begin{equation}
\label{pulse} E(t) =\exp\left(-i\omega_l t
-\frac{t^2}{4\sigma^2}(1+ia)\right) \,.
\end{equation}
We have assumed that our pulse is long enough, so one can perform
the standard decomposition of the electric field into the slow
amplitude (\ref{envelop}), and the harmonic carrier with frequency
$\omega_l$. The intensity and the power spectrum  of this pulse
are
\begin{equation}\label{intensity}
    I(t) =   \exp\left( -\frac{t^2}{2\sigma^2}\right), \quad P(\omega) =
\frac{2\sigma^2}{\sqrt{1+a^2}}\exp\left( -\frac{2\omega^2
\sigma^2}{1+a^2}\right)\,.
\end{equation}

In these formulas the full  duration of the pulse is defined as a
full width at half maximum of the intensity (FWHM): $\tau_p=
2\sqrt{2\ln 2}\,\sigma$, the linear chirp is characterized by a
real parameter $a$, and the electric field amplitude has been
conveniently selected to be one in arbitrary  units. In all
numerical applications in this paper, we select $\sigma^2=1/2$
leading to a pulse duration $2\sqrt{\ln 2}$. In order to keep our
formulas simple we have shifted the frequency $\omega$ in such a
way that it incorporates the constant carrier frequency.

The  chirp on the pulse (\ref{pulse}) corresponds to a linear
chirp $\dot{\varphi}(t)= \frac{a}{2\sigma^2}t$, leading to the
instantaneous pulse frequency $\omega(t) =-\frac{a}{2\sigma^2}t$.
For the Gaussian pulse the formula (\ref{dispuls}) becomes
\begin{equation}
  (\Delta \omega)^2=\frac{1}{4\sigma^2} +
  \frac{a^2}{4\sigma^2}\langle t^2\rangle\,.
\end{equation}
We see that  the linear chirp is equivalent to a transformation
given by the following $ABCD$ matrix
\begin{equation}
 \left(%
\begin{array}{cc}
  A & B \\
  C & D \\
\end{array}%
\right)=
\left(%
\begin{array}{cc}
  1& 0 \\
 \frac{a}{2\sigma^2} & 1 \\
\end{array}%
\right)\,.
\end{equation}
This matrix has the typical form  for a ray transformation due to
a  thin lens \cite{pmjhe}.

\begin{figure}[h]
\begin{center}
\includegraphics[scale=0.9]{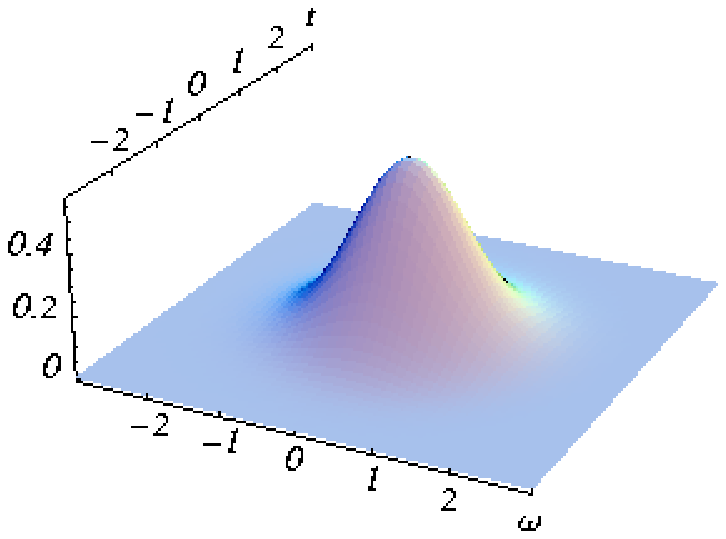}
\caption{Plot of the Wigner function for a Gaussian pulse with
pulse duration $\tau_d = 2\sqrt{\ln 2}$ and no chirp ($a=0$).}
\label{wiga0}
\end{center}
\end{figure}

\begin{figure}[h]
\begin{center}
\includegraphics[scale=.9]{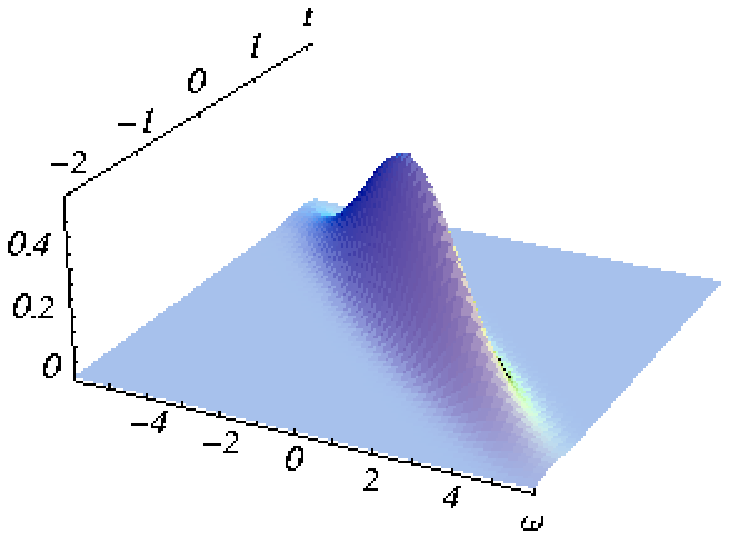}
\caption{Plot of the Wigner function for a Gaussian pulse with
pulse duration $\tau_d = 2\sqrt{\ln 2}$ and  chirp $a=3$.}
\label{wiga5}
\end{center}
\end{figure}

Let us investigate the time and frequency properties of the
chirped pulse. Simple calculation shows that the Wigner function
of this pulse is
\begin{eqnarray}\label{wigchirp}
W(t,\omega) &=& W_{0}\bigl(t,\omega+ \frac{a}{2\sigma^2}t\bigr)
\nonumber\\
&=&\frac{2\sigma}{\sqrt{2\pi}} \exp\left(
-\frac{t^2}{2\sigma^2}(1+a^2) -2\omega^2\sigma^2 -2a\omega
t\right)\,,
\end{eqnarray}
where $W_{0}(t,w) =\frac{2\sigma}{\sqrt{2\pi}} \exp\left(
-\frac{t^2}{2\sigma^2} -2\omega^2\sigma^2 \right)$ is the Wigner
function of a Gaussian pulse with no chirp. In Figures
(\ref{wiga0}) and (\ref{wiga5}) we have depicted the Wigner
function of a Gaussian pulse with no chirp and chirp $a=3$.

For the Ambiguity function a simple calculation for the chirped
pulse gives an expression similar to the formula (\ref{wigchirp})
\begin{eqnarray}
 A_{E}(T,\Omega) &=&  A_{E}(0,0) A_{0}(T,\Omega-\frac{a}{2\sigma^2}T)\nonumber\\
 &=& A_{E}(0,0)\exp\left(
-\frac{T^2}{8\sigma^2}(1+a^2) -\frac{\Omega^2\sigma^2}{2}
+\frac{a}{2}T\Omega\right)\,.
\end{eqnarray}
This function can be written  as a general Gaussian function of
two variables
\begin{equation}
A_{E}(T,\Omega) = A_{E}(0,0)\exp\left(
 -\frac{1}{2}\mathbf{v}\,
 \mathbf{C}\,
 \mathbf{v}^T\right)\,,
\end{equation}
where $\mathbf{v}= (T,\Omega)$ is a vector and $\mathbf{C}$ is a
$2\times2$ covariance matrix of the time and frequency variables.
As a result
\begin{equation}\label{covariance}
  \mathbf{C}= \left[ \begin{array}{cc}
    \langle t^2\rangle  & \langle t \omega\rangle  \\
    \langle t\omega\rangle  & \langle \omega^2\rangle  \\
  \end{array}\right]
  = \left[ \begin{array}{cc}
    \sigma^2 & -\frac{a}{2} \\
   -\frac{a}{2}  & \frac{1+a^2}{4\sigma^2} \\
  \end{array}\right]\,.
\end{equation}
Note that
\begin{equation}\label{detC}
\det \mathbf{C}= \frac{1}{4}\,.
\end{equation}
From this covariance function we obtain that time and frequency
dispersions are
\begin{equation}\label{disper}
\Delta t  =\sigma, \quad \Delta\omega =\frac{1}{2\sigma}
\sqrt{1+a^2}\,,
\end{equation}
and that the Fourier uncertainty relation between frequency
dispersion and time dispersion is
\begin{equation}\label{uncert}
    \Delta t\, \Delta \omega =\frac{ \sqrt{1+a^2}}{2} \geq
    \frac{1}{2}\,,
\end{equation}
where the lower bound corresponds to Gaussian pulses with no
chirp. From these relations we see that  the chirp enlarge the
spectral width of the pulse.

From the covariance matrix (\ref{covariance}) we see that the
chirped pulse leads to the following  time-frequency correlation
\begin{equation}\label{tfreqcorr}
   \langle t\,\omega\rangle =-\frac{a}{2}\,.
\end{equation}
In the next Section we show that   the Ambiguity function is
particularly useful to quantify the ``strength'' of this
time-frequency correlation.

\subsection{Schmidt's decomposition  of chirped pulses}

We will use the Schmidt decomposition in order to quantify the
correlation properties of the chirped pulses. This decomposition
has been successfully used  to quantify entanglement of quantum
mechanical systems described by a correlated two-party wave
functions. As an example we note that recently Professor Eberly
has investigated high transverse entanglement in optical
parametric down conversion using a Schmidt decomposition  of the
biphoton wave function \cite{laweberly2004}.

In order to apply the Schmidt decomposition we replace
$(T,\Omega)$ by two dimensionless variables $(X,Y)$ such that the
Ambiguity function takes the following form
\begin{equation}
\label{AXY}
 A(X,Y) = A(0,0) \exp( -X^2-Y^2 -2 cXY)\,,
\end{equation}
where
\begin{equation}\label{cnumber}
    c= \frac{\langle t \omega\rangle}
    {\sqrt{\langle t^2 \rangle\langle  \omega^2\rangle }
    }\,.
\end{equation}
The Schmidt decomposition of the  function (\ref{ambfunc}) is

\begin{equation}\label{schmidt}
 A(X,Y) = A(0,0)  \sum_n \sqrt{p_n}\; \psi_{n}(X)\psi_{n}(Y)
\end{equation}
where $\psi_{n}(X)$ and $\psi_{n}(Y)$ are the Schmidt modes
defined as eigenstates of the reduced density operators
constructed from a two-party wave function given by the Ambiguity
function (\ref{AXY}). The Schmidt eigenvalues $p_n$ serve as a
degree of entanglement or correlation between the two-part system.
In our case the Schmidt eigenvalues quantify the degree of
correlation between time and frequency. The measure of correlation
can be quantified by entropy of entanglement
\begin{equation}\label{entropy}
E=  1- \sum_n p_n^2  \quad \Rightarrow \quad  0\leq E \leq 1\,.
\end{equation}
For a untangled pulse we have  $E=0$ because there is   only one
Schmidt eigenvalue, and  the formula  (\ref{schmidt}) factorizes:
$A(X,Y)=A(X)A(Y)$.

 The Schmidt decomposition can be performed for a  Gaussian ambiguity
function characterized by an arbitrary covariance matrix
(\ref{covariance}). In this case the entropy of entanglement is
\begin{equation}
\label{coentropy}
 E = 1- \sqrt{1- \frac{\langle t \omega\rangle^2}{
\langle t^2 \rangle \langle  \omega^2\rangle} } \,.
\end{equation}
Clearly we have $E=0$ for uncorrelated in time and frequency
optical pulses.

\begin{figure}[h]
\begin{center}
\includegraphics[scale=0.9,angle=0]{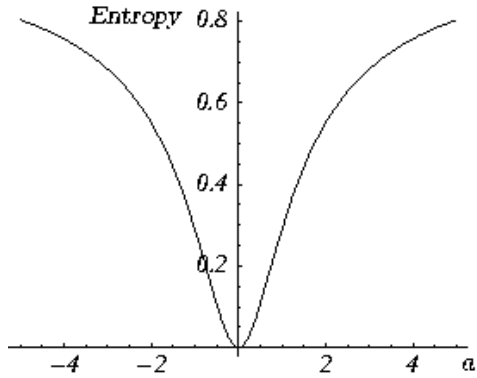}
\caption{Entropy of the pulse as a function of the chirp $a$
}\label{entr}
\end{center}
\end{figure}

For the Gaussian optical pulse (\ref{pulse}) with chirp, we obtain
that  the Schmidt eigenvalues have the formal form of the
Bose-Einstein distribution
\begin{equation}\label{pn}
p_n = \frac{\bar{n}^n}{(1+\bar{n})^{n+1}}, \quad \mathrm{with}
\quad \bar{n}= \frac{1}{2}(\sqrt{1+a^2} -1)\,.
\end{equation}
The Schmidt eigenstates corresponds to a thermal distribution with
a mean number of photons given by $\bar{n}$. Using this expression
we calculate  the entropy of entanglement (\ref{coentropy})
\begin{equation}
\label{entropypulse}
E = 1- \frac{1}{\sqrt{1+a^2}}\,.
\end{equation}
In Figure (\ref{entr}), we have depicted the  entropy of
entanglement as a function of the chirp $a$.

We conclude this section with a remark that pulses with linear
chirp are analogous to quantum mechanical wave functions of the
form $\psi(x)= e^{iax^2}$. In the framework of quantum mechanics
such wave functions belong to a wide class of the so-called
contractive states \cite{kw1984b}, and  have been used to show the
narrowing of the position uncertainty relation \cite{yuen1983},
and applied by Professor Eberly to phase entanglement
\cite{kcjhe2004}. The contractive nature of the chirped pulse can
be easily exhibited using an lens type $ABCD$ transformation of
the chirped pulse, equivalent to the following linear
transformation  in time and frequency plane
\begin{equation}
\omega_t=\omega+\frac{t}{2\sigma^2}\,.
\end{equation}
For a chirped pulse the dispersion of this observable is
\begin{equation}
(\Delta \omega_t)^2= (\Delta \omega)^2 +
\frac{1}{4\sigma^4}(\Delta t)^2
    +\frac{1}{\sigma^2}\langle \omega t\rangle
    = \frac{2+a^2}{\sigma^2} - \frac{a}{2\sigma^2}\,.
\end{equation}
For positive $a$, this formula exhibits a contraction of the
uncertainty of $\omega_t$ similar to the narrowing of a freely
evolving quantum mechanical wave packets. The narrowing is
entirely due to the time-frequency correlation of the chirped
pulse (\ref{tfreqcorr}).

\subsection{Time dependent spectrum for chirped pulses}

In order to calculate the time-dependent spectrum for the chirped
pulse, we will  use for the filter response
\begin{equation}
\mathcal{J}(t) = \theta(t) \Gamma \sqrt{\frac{2}{\pi}}\exp\left(
-\frac{t^2}{4\Gamma^2}\right)\,.
\end{equation}
This function corresponds to a linear, causal and time translation
invariant response of the filter. In the calculations  we will
assume that the observation time is larger compared with the pulse
duration: $t\geq \tau_d$.  Simple calculation shows that the
spectrum is Gaussian and has the form
\begin{equation}\label{tdschirp}
G(t,\omega) =G_0 \exp\left(
 -\frac{1}{2}\mathbf{w}\,
 \mathbf{G}\,
 \mathbf{w}^T\right)
\end{equation}
where $\mathbf{v}= (t,\omega)$ is a vector and the matrix
$\mathbf{G}$ can be expressed in terms of the covariance matrices
of the pulse and the filter
\begin{equation}
 \mathbf{G}^{-1}= \mathbf{C}+ \mathbf{C}_{\mathcal{J}}=
\left[ \begin{array}{cc}
   \Gamma^2+ \sigma^2 & -\frac{a}{2} \\
   -\frac{a}{2}  & \frac{1}{\Gamma^2}+ \frac{1+a^2}{4\sigma^2} \\
  \end{array}\right]\,.
  \end{equation}
The normalization constant $G_0$ of the time-dependent spectrum
is such that
\begin{equation}
\int \frac{d t d\omega}{2\pi}\, G(t,\omega)= \int dt\, I(t)\,.
\end{equation}

\section{Linear superposition of chirped pulses}

Linear superposition principle play a fundamental role in
classical and quantum interference phenomena. In order to
illustrate time and frequency interference we shall investigate a
linear superposition of two optical pulses
\begin{equation}\label{linsuper}
 E_{\mathrm{sup}} = E(t-t_0) + E(t+t_0)\,,
\end{equation}
where $2t_0$ is a temporal separation between the  pulses. This
linear superposition of two electric fields of optical pulses
exhibit classical interference very similar to the interference of
quantum coherent states \cite{janszky1994,kwgh1998}.

\begin{figure}[h]
\begin{center}
\includegraphics[scale=0.9,angle=0]{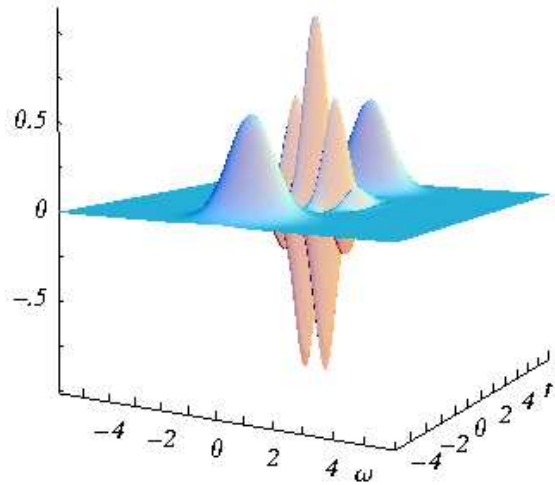}
\caption{Plot of the Wigner function for two Gaussian pulses with
no chirp $a=0$ and $t_0=4$.}\label{2wig05}
\end{center}
\end{figure}

\begin{figure}[h]
\begin{center}
\includegraphics[scale=0.9,angle=0]{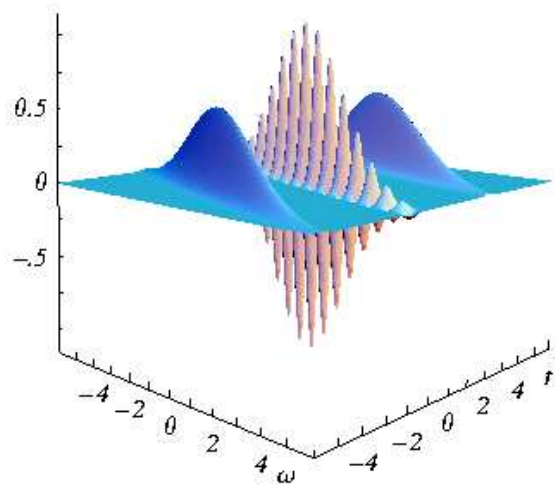}
\caption{Plot of the Wigner function for two Gaussian pulses with
 chirp $a=3$ and $t_0=4$.}\label{2wig55}
\end{center}
\end{figure}

As an example of such a superposition we take two Gaussian chirped
pulses
\begin{equation}
\label{2pulse} E_{\mathrm{sup}}(t) =\exp\left(
-\frac{(t-t_{0})^2}{4\sigma^2}(1+ia)\right) + \exp\left(
-\frac{(t+t_{0})^2}{4\sigma^2}(1+ia)\right)\,.
\end{equation}
The intensity of this superposition is:
\begin{equation}\label{2intensity2}
 I(t) = \exp\left( -\frac{(t-t_{0})^2}{2\sigma^2}\right)+ \exp\left(
 -\frac{(t+t_{0})^2}{2\sigma^2}\right) +2 \exp\left(
 -\frac{t^2+t_{0}^2}{2\sigma^2}\right)\cos\left(\frac{tt_{0}a}{\sigma^2}\right)\,.
\end{equation}
The corresponding power spectrum is
\begin{equation}\label{2power}
P(\omega)=
\frac{4\sigma^2}{\sqrt{1+a^2}}\exp\left(-\frac{2\omega^2
\sigma^2}{1+a^2}\right) \cos^2(\omega t_0)\,.
\end{equation}
The time-frequency Wigner function of the linear superposition
(\ref{2pulse}) is
\begin{eqnarray}\label{2wigchirp}
W(t,\omega) &=& W_{E}(t+t_0,\omega)+ W_{E}(t- t_0,\omega)
+2 W_{E}(t,\omega)\cos(2t_0\omega)\,.
\end{eqnarray}
In this formula the Wigner function: $W_{E}(t,\omega)$ is given by
(\ref{wigchirp}). From the Wigner function of the superposition
it is possible to calculate time and frequency moments. Simple
calculation leads to
\begin{equation}\label{2tdispdef}
   \langle t^2\rangle =\sigma^2 + \frac{t_0^2(1 -a^2\exp(
   -\frac{t_0^2(1+a^2)}{2\sigma^2}))}{1+\exp(
   -\frac{t_0^2(1+a^2)}{2\sigma^2})}\,,
\end{equation}
and
\begin{equation}\label{2fredispdef}
   \langle \omega^2\rangle
   =\frac{1+a^2}{4\sigma^2}\left(1-\frac{t_0^2}{\sigma^2}(1+a^2)\frac{\exp(
   -\frac{t_0^2(1+a^2)}{2\sigma^2})}{1+\exp(
   -\frac{t_0^2(1+a^2)}{2\sigma^2})}\right)\,.
\end{equation}
From this formula we see that the spectrum of the linear
superposition is reduced ({\it squeezed}) below the single pulse
width. This effect is entirely due to the fact that we are dealing
with a linear superposition  of pulses. In Figures (\ref{2wig05})
and (\ref{2wig55}) we have depicted the Wigner function for the
linear superposition with $t_0=4$ with no chirp and  chirp $a=3$.
The squeezing effect corresponding to the nonzero chirp is clearly
seen.

Note that in reference \cite{beck1993}, the Wigner function of a
coherent two-pulse sequence with linear frequency chirp has been
reconstructed experimentally using quantum tomography.

\begin{figure}[h]
\begin{center}
\includegraphics[scale=0.9,angle=0]{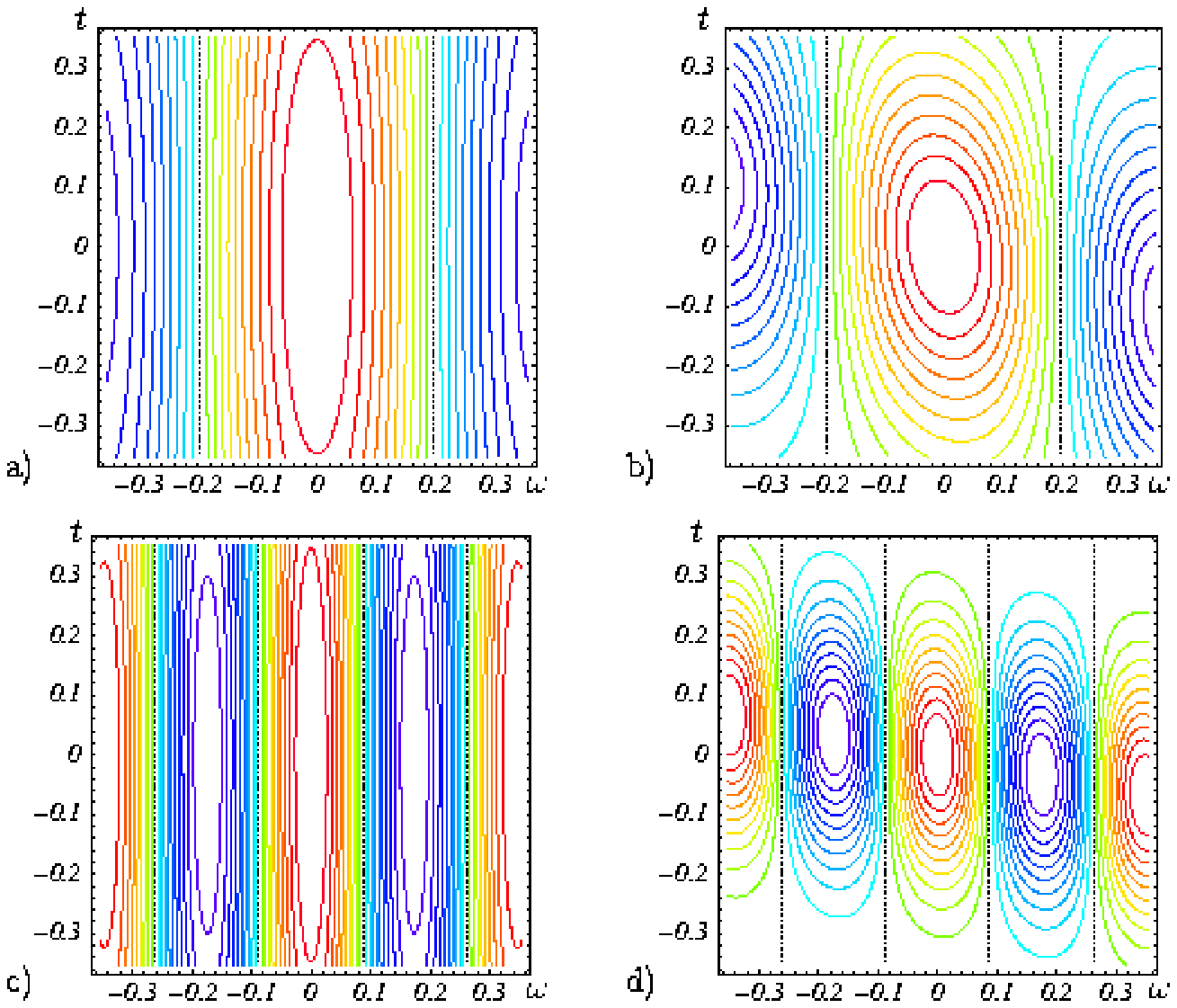}
\caption{The interference terms of the Wigner function for two
Gaussian pulses plotted at the region of the phase space below the
Fourier uncertainty: a)  fragment of the plot from Figure
(\ref{2wig05}), i.e. $a=0$, $t_0=4$; b) fragment of the plot from
Figure (\ref{2wig55}), i.e.  $a=3$, $t_0=4$; c), d) similar terms
obtained for larger time separation distance $t_0=9$ with chirps
$a=0$ and $a=5$, respectively.} \label{sub}
\end{center}
\end{figure}

The remarkable feature of the Wigner function (\ref{2wigchirp}) is
the fact that it contains structures in a phase space region below
the Fourier uncertainty relation. In Figure (\ref{sub}) we have
depicted the Wigner function in a space region with $\Delta\omega
\leq 1/\sqrt{2}$ and $\Delta t \leq 1/\sqrt{2}$. In the framework
of quantum mechanics it has been recognized that  small structures
on the  sub-Planck scale do show up in quantum linear
superpositions \cite{zurek2001}. It is clear that for linear
superpositions  of chirped pulses such sub-Fourier structures
emerge as well. We will see in the next Section, that due to such
small structures it will be possible to have pulses with zero
overlap.
%

\section{Time and frequency overlap}
\begin{figure}[h]
\begin{center}
\includegraphics[scale=0.9,angle=0]{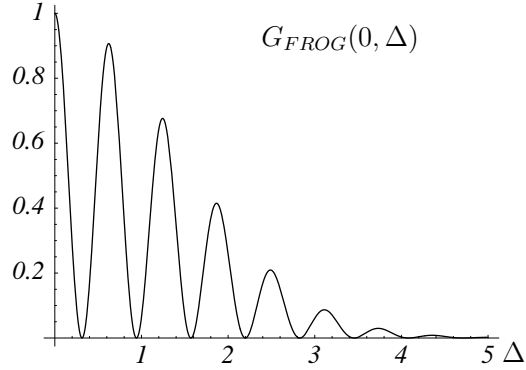}
\caption{The overlap as a function of $\Delta$  for a linear
superposition with $t_0=5$ and $a=5$ }\label{over}
\end{center}
\end{figure}

In this Section we will investigate a FROG version (with
$\mathcal{J}(t)= E^{\ast}_{\mathrm{sup}}(t)$) of the time
dependent spectrum (\ref{tds}) applied to the linear
superposition of pulses given by the formula (\ref{2pulse})
\begin{equation}
 G(0,\Delta)=\biggl|\int ds\,
 E^{\ast}_{\mathrm{sup}}(s)E_{\mathrm{sup}}(s)e^{i\Delta
 s}\biggl|^2\,.
\end{equation}
Such an overlap  can be easily obtained if one of the two pulses
has its  carrier frequency detuned by $\Delta $. In this case the
FROG signal is a phase space overlap (\ref{overlap}) of the form
\begin{equation}\label{overlap1}
G_{FROG}(0,\Delta)=  2\pi \int dt \int d\omega\,
W_{E_{\mathrm{sup}}}(t,\omega)
W_{E_{\mathrm{sup}}}(t,\omega+\Delta)\,.
\end{equation}
For the chirped pulses this overlap can be calculated, and as a
result we obtain
\begin{equation}\label{overlap2}
G_{FROG}(0,\Delta)= \mathcal{N}
e^{-\frac{\Delta^2\sigma^2}{2}}\left( \cos(\Delta t_0) +e^{
-\frac{t_{0}^2}{2\sigma^2}(1+a^2)}\cosh(\Delta t_0 a)\right)^2\,,
\end{equation}
where $\mathcal{N}$ is a normalization constant. In Figure
(\ref{over}) we have depicted (\ref{overlap2}) as a function of
$\Delta$, for $t_0=5$ and $a=5$.

\begin{figure}[h]
\begin{center}
\includegraphics[scale=0.9,angle=0]{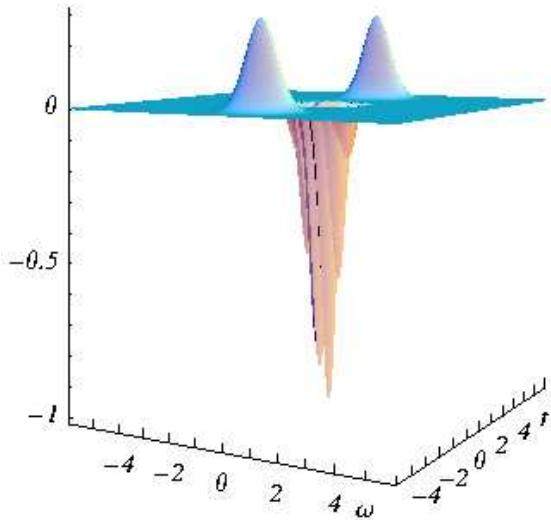}
\caption{The product of two Wigner functions for linear
superpositions of pulses with $t_0=4$ and $a=0$ mutually shifted
in carrier frequencies by $\Delta=\frac{\pi}{2t_0}$.
}\label{Wig_product}
\end{center}
\end{figure}

The remarkable feature of this curve is that $G_{FROG}(0,\Delta)$
can vanish for values of $\Delta$ smaller then the frequency
uncertainty in frequency. In terms of the time and frequency
distribution function, the detuning $\Delta$  of the carrier
frequency corresponds just to a shift of the Wigner function along
the $\omega$ axis. In the case of the superposition of two
Gaussian pulses, this shift affects the Gaussian peaks and  the
oscillating non-positive interference term. It is easy to notice
that for an appropriate value of $\Delta$ this can result  in a
sign change  of the interference term in comparison to the
original function. Calculating  the overlap (\ref{overlap1}) we
need a product of shifted and unshifted Wigner functions. For a
selected value of the shift  $\Delta$, the product of the two
interference terms of their Wigner functions will become negative
(or zero). In   Figure (\ref{Wig_product}) we have depicted such a
product with $t_0=4$ and shift $\Delta=\pi/2t_0$. We see that the
considerably large negative contribution can cancel the positive
peaks corresponding to the overlap of the non-interference terms.
Obviously, formula (\ref{overlap2}) can be easily
obtained without using the Wigner function, but the phase-space
representation (\ref{overlap1}) is especially useful to  show that
the sub-Fourier structures are peculiar to interference phenomena.

In quantum mechanical framework this feature is especially
interesting as the zero--overlap means that the corresponding
states are orthogonal, and so at a sub-Planck scale one can obtain
sets of mutually orthogonal states \cite{zurek2001}. In the case
of optical pulses we can obtain a similar result of  pulses with
zero--overlap using a shifts  of frequency that are below the
Fourier uncertainty relation. Certainly, the smaller is the value
of this shift $\Delta$ that one wants to use, the larger $t_0$ has
to be taken in the calculations, meaning that the
Fourier/Heisenberg uncertainty relation is satisfied.

\section{Conclusions}
In this paper we have analyzed the connection of the time and
frequency time-dependent spectrum with phase space distributions
based on the Wigner and the Ambiguity functions. We have exploited
the similarities between optical pulses and quantum mechanical
wave packets to  exhibit  interference, entanglement,  and
 sub-Fourier structures of the time and
frequency phase space.

Certainly the Wigner function is not the only possible way to give
a time and frequency description of optical pulses. Recently in
reference \cite{lpkw2003} we have introduced a class of new phase
space representations based on the so called Kirkwood--Rihaczek
distribution function \cite{cohen1995}:
\begin{equation}
  K(t,\omega)
 =\int \frac{d s}{2\pi} \; E^{\ast}(s)
\; e^{i\omega (s-t)} \; E(t)\,.
\end{equation}
Such phase space distributions provide  time and frequency
characteristics of optical pulses, but only phase space overlaps
of these distributions have an operational meaning as a physical
spectrum of light.  In the case of the Kirkwood--Rihaczek function
this overlap,  much resembling the equation (\ref{overlap}), is
given by the following formula
\begin{equation}\label{overlapKR}
 \left|\int dt E^{\ast}_{1}(t)E_{2}(t)\right|^2=
2\pi \int dt \int d\omega\, K_{E_1}(t,\omega) K^*_{E_2}{(}t,\omega)\,.
\end{equation}
Thus, for the superpositions discussed in Section V, we can write
the following formula
\begin{equation}
G_{FROG}(0,\Delta)=  2\pi \int dt \int d\omega\,
K_{E_{\mathrm{sup}}}(t,\omega)
K^{*}_{E_{\mathrm{sup}}}(t,\omega+\Delta)\,.
\end{equation}

We conclude this paper by noting that the  definition of the
time-dependent spectrum of light introduced 28 years ago, is still
an attractive field of research producing interesting insights
into the definition of the spectrum of nonstationary ensemble of
optical pulses \cite{wolf2004}.

\section*{Acknowledgments}
This article is dedicated to Professor J. H. Eberly. We honor
Professor Eberly contributions to the development of the
time-dependent spectrum and to the broad field of Quantum Optics.
This work was partially supported by the Polish Ministry of Scientific 
Research Grant 
PBZ-Min-008/P03/03. 
KW thanks J-C. Diels and G. Herling for interesting comments related to
chirped pulses.


\end{document}